\documentclass{article}
\begin{document}
\def\p {{\partial}}
\def\n {{\nu}}
\def\m {{\mu}}
\def\a {{\alpha}}
\def\bt {{\beta}}
\def\f {{\phi}}
\def\th {{\theta}}
\def\g {{\gamma}}
\def\eps {{\epsilon}}
\def\e {{\psi}}
\def\la {{\lambda}}
\def\na {{\nabla}}
\def\bn {\begin{eqnarray}}
\def\en {\end{eqnarray}}
\title{Path integral formulation of constrained systems with
singular-
higher order Lagrangians}
\maketitle
\begin{center}
\author{S.I. MUSLIH\\Dept. of Physics\\ Al-Azhar university\\
Gaza, Palestine}
\end{center}

\begin{abstract}
Systems with singular higher order- Lagrangian are investigated
by using the extended form of the canonical method. Besides, the
canonical path integral formulation is generalized using the
Hamilton - Jacobi formulation to investigate singular systems.
\end{abstract}
\section{Introduction}

In spite of the fact that most physical systems can be described
by Lagrangians that depend at most on the first derivatives of
the dynamical variables [1-4] there is a continuing interest in
the so called generalized dynamics, that is, the study of
physical systems described by Lagrangians containing derivatives
of order higher than the first.

The generalization of Hamilton's least action principle and of
the Hamiltonian formulation to non degenerate Lagrangian
depending on higher-order derivatives was first achieved by
Ostrogradisky. [5].

Recently a new method [6, 7] based on the Hamilton- Jacobi method
[8-11] has been developed to investigate singular systems. The
aim of this paper is to study the path integral quantization for
singular systems with arbitrarily higher order-Lagrangian. In fact
this work is a continuation of previous papers [6,7], where we
have obtained the path integral for singular systems with first
order Lagrangians. Our desire is to construct the path integral
quantization for singular systems starting from the Hamilton
-Jacobi Partial differential equations. HJPDE, which is the
fundamental equation of classical mechanics. In this method the
equations of motion are obtained as total differential equations
in many variables which require the investigation of
integrability conditions. If the system is integrable, one can
construct the canonical phase space and the canonical action is
obtained by this procedure. Hence, one can obtain the path
integral formulation as an integration over the canonical phase
space coordinates
\section{The extended canonical
path integral method}

Now we will construct the canonical path integral by using the
Hamilton- Jacobi method [8-11]. The starting point of this
procedure is to consider a system described by a Lagrangian

\bn &&L(q_1,...,q_i^{(k)}),  q_i^{(l)}=\frac{d^l q_i}{dt^l},
\nonumber\\&&l=0,1,...,k-1,\; i=1,...,n, \en where the
derivatives $q_i^{(s)}(s=0,1,...,k-1)$ are treated as coordinates.
In Ostogrodoski's formula the momenta conjugated respectively
to $q_{(k-1)i}$ and $q_{(m-1)i}(m=1,...,k-1)$ are defined as
\begin{equation}
p_{(k-1)i}=\frac{\p L}{\p q_i^{(k)}},
\end{equation}
\begin{equation}
p_{(m-1)i}=\frac{\p L}{\p q_i^{(m)}}-p_{(m)i},\;m=1,2,...,k-1,
\end{equation}
using these relations one can go over from the Lagrangian
description to the Hamiltonian description. The canonical
Hamiltonian is defined as
\begin{equation}
H_{0}= \sum _{s=0}^{k-1}p_{(s)i}q_i^{(s+1)}- L(q_1,...,q_i^{(k)},
q_i^{(l)}).
\end{equation}

"Einstein's summation rule for repeated indices is used
throughout this paper".

Now the extended Hessian matrix is
defined as
\begin{equation}
A_{ij}= \frac{\p^2 L}{\p q_i^{(k)}\p q_j^{(k)}},
\end{equation}
For a regular system, the Hessian has rank $n$ and the canonical
coordinates are independent. For singular Lagrangian case the
Hessian has rank$n-r$, $r < n$. In this case $r$ of the momenta
are dependent. The generalized coordinate $q_{(k-1)i}$ are
defined as
\begin{equation}
p_{(k-1)a}=\frac{\p L}{\p q_a^{(k)}},\;a=r+1,...,n,
\end{equation}
\begin{equation}
p_{(k-1)\m}=\frac{\p L}{\p q_\m^{(k)}},\;\m=1,...,r.
\end{equation}

Since the rank of the Hessian is $(n-r)$ one may solve equation
(6) for $q_{(k-1)\m}$ as functions of $t$,
$~q_{(s)i}$,$p_{(k-1)b}$and $ q_{\m}^{(k)}$as follows
\begin{equation}
q_a^{(k)}=W_{(k)a}(q_{(s)i},\;p_{(k-1)b},\; q_{\m}^{(k)}),\;
b=r+1,...,n.
\end{equation}
Now substituting equation (8) in equation (7)
one has
\begin{equation}
p_{(k-1)\m}=\frac{\p L}{\p
q_\m^{(k)}}\mid_{q_a^{(k)}=W_{(k)a}(q_{(s)i},p_{(k-1)b},
q_{\m}^{(k)}) },
\end{equation}
or

\bn &&p_{(s)\m}= -H_{(s)\mu}(t,q_{(u)j}; p_{(u)a}=\frac{\p S}{\p
q_{(u)a}}),\nonumber\\ &&u,s=0,...,k-1,j=1,...,n. \en

Relapling the coordinates $t$ as $t_{(s)0}\equiv q_{(s)0}$ (for
any value of$s$); the coordinates $ q_{(s)\m}$ will be called
$q_{(s)\m}$ and defining $p_{(s)0}=\frac{\p S}{\p t}$, while
$H_{(s)0}=H_{0}$ for any value of $s$. In this case the canonical
Hamiltonian $H_{0}$ may be written as \bn
&&H_{0}= \sum_{u=0}^{k-2}p_{(u)a}q_a^{(u+1)}+p_{(k-1)a}W_{(k)a}+\sum
_{u=0}^{k-1}q_\m^{(u+1)}p_{(u)\m}{\mid}_{p_{(s)\n}=H_{(s)\n}}\nonumber\\&&-
L(q_{(s)i}1,...,q_{\m}^{(k)},
q_a^{(k)}=W_{(k)a}),\nonumber\\&&\m,~\n=1,...,r,a=r+1,...,n. \en

Now the canonical method leads to obtain the set of
Hamilton-Jacobi partial differential equations as follows

\bn &&H^{'}_{0}=H^{'}_{(s)0}=p_{(s)0}+ H_{(s)0}(t,t_{(u)\m};
q_{(u)a},p_{(u)a}=\frac{\p S}{\p q_{(u)a}})=0,\\
&&H^{'}_{(s)\m}=p_{(s)\m}+ H_{(s)\m}(t_{(u)\n};
q_{(u)a},p_{(u)a}=\frac{\p S}{\p
q_{(u)a}})=0,\nonumber\\&&u,s=0,...,k-1,,\m,\n==1,...,r,\en
or

\bn &&H^{'}_{(s)\a}=p_{(s)\a}+ H_{(s)\a}(t_{(u)\bt};
q_{(u)a},p_{(u)a}=\frac{\p S}{\p q_{(u)a}}),\nonumber\\
&&\a,\bt=0,1,...,r. \en

The equations of motion are obtained as total differential
equations in many variables as follows:

\bn
dq_{(u)i}=&& \sum
_{s=0}^{k-1}\frac{\p H^{'}_{(s)\a}}{\p p_{(u)i}}dt_{(s)\a};\\
&&i=1,...,n,\a=0,1,...,r,u,s=0,1,...,k-1.\nonumber\\
dp_{(u)c}=&& -\sum
_{s=0}^{k-1}\frac{\p H^{'}_{(s)\a}}{\p q_{(u)i}}dt_{(s)\a};\\
&&c=0,1,...,r,\a=0,1,...,r,u,s=0,1,...,k-1.\nonumber\\
dZ =&&\sum _{d=0}^{k-1}[-H_{(d)\bt}+  \sum
_{s=0}^{k-1}p_{(s)a}\frac{\p
H^{'}_{(d)\bt}}{\p p_{(s)a}}]dt_{(d)\bt};\\
&&s,d=0,...,k-1,\bt=0,1,...,r,\nonumber \en
where
$Z=S(t_{(u)\a};q_{(s)\a})$. The set of equations (15-17) is
integrable [8] if
\begin{equation}
dH^{'}_{(s)\a}(t_{(u)\n}; q_{(u)a};p_{(u)\n}=\frac{\p
S}{\p_{(u)\n}}; p_{(u)a}=\frac{\p S}{\p q_{(u)a}})=0,\a=0,1,...,r,
\end{equation}
conditions (18)considering equations (15-17), may vanish
identically or give rise to new constraints. In the case of new
constraints one should consider their total variations also.
Repeating this procedure one may obtain a set of conditions such
that all the total variations vanish. Simultaneous solutions of
canonical equations with all these constraints provide the
solutions of a singular system.$H^{'}_{(s)\a}$ can be interpreted
as infinitesimal generators of canonical transformations given by
parameters $t_{(s)\a}$ respectively. In this case as for the
first-order systems, the path integral may be written as

\bn
D({q'}_i^{(u)},{t'}_{(u)\a};q_i^{(u)},t_{(u)\a})=&&\int_{q_i^{(u)}}^{{q'}_i^{(u)}}~~~~~~~dq_{(u)a}~dp_{(u)a}\times
\nonumber\\&& \exp i\{\int_{t_{(u)\a}}^{{t'}_{(u)\a}} \sum
_{d=0}^{k-1}[-H_{(d)\bt}+  \sum _{s=0}^{k-1}p_{(s)a}\frac{\p
H^{'}_{(d)\bt}}{\p
p_{(s)a}}]dt_{(d)\bt}\},\nonumber\\&&u,s,d=0,1,...,k-1,\a,\bt=0,1,...,r,
\nonumber\\&&a=r+1,...,n. \en The path integral expression (19)
is an integration over the canonical phase space
coordinates$(q_{(u)a},p_{(u)a})$.

\section{ Conclusion}

We have obtained the canonical path integral formulation of
singular higher-order systems. In this formulation, the equations
of motion are obtained as total differential equations in many
variables which require the investigation of integrability
conditions (18). If the system is integrable then each coordinate
$q_{(s)\a}= t_{(s)\a}(\a=1,...,r)$ is treated as a parameter that
describes the system evolution. The Hamiltonian $H^{'}_{(s)\a}$
will be the infinitesimal generators of canonical transformations
given by parameters $t_{(s)\a}$respectively in the same way the
Hamiltonian $H_{0}$ is the generator of time evolution. For $k=1$
the result obtained here (equation(19)) will reduce the case of
the first order path integral showed in references [6,7].

\end{document}